\RequirePackage{fix-cm}
\documentclass[smallextended]{svjour3}       
\smartqed  
\usepackage{graphicx}
\usepackage{amssymb}

\usepackage[dvipsnames]{xcolor} 

\begin{document}

\title{Simulated observations of heavy elements with CUBES}


\titlerunning{Heavy element abundances with CUBES}        

\author{H. Ernandes$^1$, B. Barbuy$^1$, B. Castilho$^2$, C. J. Evans$^3$, 
G. Cescutti$^4$
        
}

\authorrunning{H. Ernandes et al.} 

\institute{H. Ernandes, B. Barbuy, \at
Universidade de S\~ao Paulo, IAG, Rua do Mat\~ao 1226, S\~ao Paulo, 05508-900, Brazil \\
             \email{heitor.ernandes@usp.br}
          \and
          B. Castilho \at
LNA/MCTI, Rua Estados Unidos, 154 - 37504-364, Itajub\'a, Brazil
\and
C. J. Evans \at
UK ATC, Royal Observatory, Blackford Hill, Edinburgh, EH9 3HJ, UK
\and
G. Cescutti \at
INAF - Osservatorio Astronomico di Trieste, Via Tiepolo 11, I-34131 Trieste, Italy}

\date{Received: date / Accepted: date}

\maketitle

\begin{abstract}
We investigate the feasibility of robust abundances for selected neutron-capture elements (Ge, Bi, Hf, U) from near-UV spectroscopy with the CUBES instrument now in development for the Very Large Telescope. We use the CUBES end-to-end simulator to synthesise observations of the Ge~I 3039\,\AA\ and Hf~II 3400 and 3719\,\AA\ lines in a very metal-poor star, using the well-studied star CS~31082-001 as a template. From simulated 4\,hr exposures, we recover estimated abundances to $\pm$0.1\,dex for Ge for $U$\,$\sim$\,14.25\,mag., and for Hf for $U$\,$=$\,18\,mag. These performances neatly highlight the powerful gain of CUBES for near-UV observations of targets that are two-to-three magnitudes fainter than the existing observations of CS~31082-001 ($U$\,$=$\,12.5\,mag.). We also investigate the weak Bi~I 3025\,\AA\ and U~II 3860\,\AA\, lines (for $U$\,$\sim$\,14.25 and 16\,mag., respectively), finding that simulated 4\,hr exposures should provide upper limits to these observationally challenging lines 

\end{abstract}

\section{Introduction}
\label{intro}
Ground-based ultraviolet (UV) spectroscopy is a treasure trove for stellar astrophysics. This part of the near-UV domain (300-400\,nm) is rich with diagnostic lines from both light- and heavy-elements, from which precision abundances can be determined for a broad range of stars.  Our observational efforts in this region are currently limited by the efficiencies of instruments such as UVES \cite{dekker00} on the Very Large Telescope (VLT). UVES has an end-to-end efficiency of only a few percent over the near-UV range.  As such, detailed chemical abundances from lines in this region for metal-poor stars, have generally been limited to relatively bright (10$^{\rm th}$ or 11$^{\rm th}$ magnitude) targets, or have required significant integration times, e.g. the 20\,hr total exposure of BPS~BS~16968-061 ($V$\,$=$\,13.2\,mag) \cite{smiljanic21}.

This is one of the scientific topics that has motivated the design of the new Cassegrain U-Band Efficient Spectrograph (CUBES), with the Phase A design of the instrument outlined elsewhere in this volume \cite{zanutta22}. CUBES will enter construction in early 2022, with science operations expected in 2028.  It will provide a continuous near-UV spectrum over the 300-405\,nm range, with the light split into two arms using a dichroic filter. The spectral resolving power of both the `blue' and `red' arms ranges from $R$\,$=$\,22,000 to 27,000 across their wavelength coverage (300-352 and 346-405\,nm, respectively), with a resolution element sampled by a minimum of 2.3 pixels on the detectors.

At the start of the Phase A study we undertook a qualitative study of the detectability of a broad range of elements in the near-UV at the spectral resolution of CUBES \cite{ernandes20}, concluding that the lower resolution (compared to UVES) was acceptable for most lines, provided there is sufficient signal-to-noise (S/N). To investigate the performance of CUBES for heavy-element abundances in a more quantitative approach, here we present simulations undertaken using the metal-poor ([Fe/H]\,$=$\,$-$2.9), r-process-enriched star CS~31082-001 as a spectral template. This star was observed extensively with UVES within the `First Stars' Large Programme at the European Southern Observatory (ESO), as well as with the {\em Hubble Space Telescope} ({\em HST}). The U/Th abundance ratio derived from the optical spectra was used to estimate the age of the star \cite{cayrel01}, and a series of papers then presented its light- and heavy-element abundances from the ground-based spectroscopy \cite{hill02,cayrel04,plez04,spite05} and the {\em HST} data \cite{barbuy11,siqueira-mello13}.  

We have revisited the UVES data of CS~31082-001 in a new study that investigated the near-UV diagnostics for several light elements compared to published values, as well as presenting the first abundances for Be, Cu and V \cite{ernandes22}. Here we present simulations investigating the detectability of selected heavy elements at the lower spectral resolving power of CUBES compared to that of UVES. In particular, we consider lines from Ge and Bi that are at the shortward end of the ground-UV domain, and which helped to motivate the blueward coverage of the instrument design.  

In Section~2 we briefly outline the elements considered in this study, followed by a description of our methods in Section~3, and our results in Sections~4 and 5. In Section~6 we close with a brief summary of the prospects for stellar spectroscopy of metal-poor stars with CUBES.



\section{Elements selected for study}

\begin{figure}
    \centering
    \includegraphics[width=0.9\linewidth]{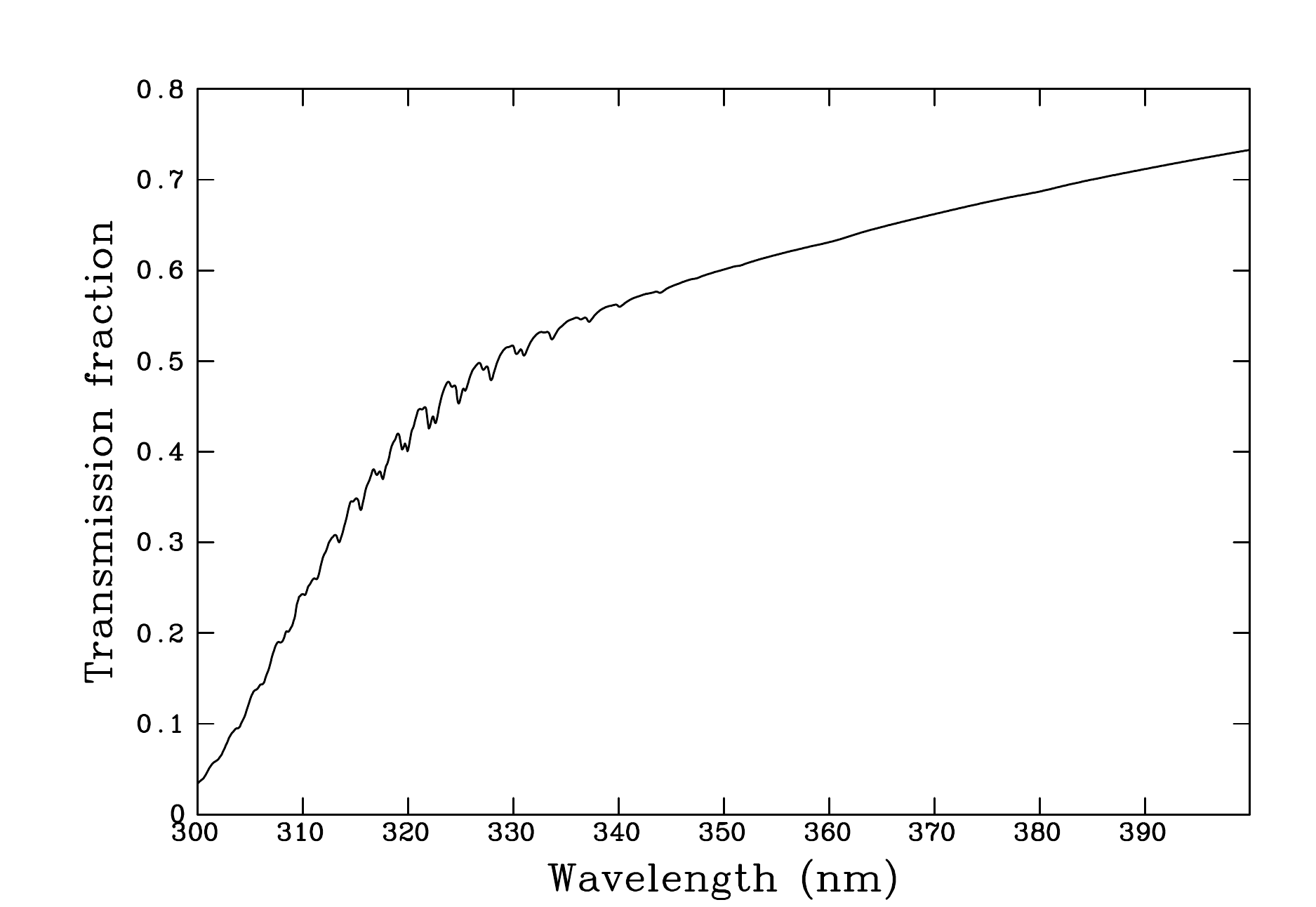}
    \caption{Mean annual sky transmission at zenith over the 300-400\,nm range for Cerro Paranal, calculated using the ESO {\tt SKYCALC} software.}
    \label{trans}
\end{figure}

The transmission of Earth's atmosphere reduces fast as we move to shorter wavelengths. Although this is effective in protecting us from damaging UV rays, it makes ground-based astronomical observations increasingly difficult shortwards of $\sim$330\,nm.  A simple illustration of the scale of this is shown in Fig.~\ref{trans}, taking data from the ESO {\tt SKYCALC} software\footnote{http://www.eso.org/observing/etc/skycalc/skycalc.htm} and calculating the sky
transmission curve over the 300-400\,nm range at zenith (averaged over the year and full nights for the Cerro Paranal site).  

With the atmospheric transmission in mind, we were strongly motivated during the Phase A study to investigate the potential performance of CUBES at the shortest possible wavelengths. As such, we initially investigate two elements (Ge, Bi) that have diagnostic lines shortwards of 305\,nm, providing motivation for the technical design to extend as far bluewards as possible \cite{zanutta22}. We also investigated two other neutron-capture elements of interest in studies of metal-poor stars (Hf, U) to further characterise the performance of the CUBES design. We now briefly summarise some of the key points for our selected elements:

\begin{itemize}
    \item {\it Germanium (Z\,$=$\,32):} Ge is often considered as either the heaviest of the iron-peak elements, or as the lightest of the trans-iron, neutron-capture elements (e.g. \cite{peterson20}). Critically, in the context of the CUBES design requirements, the only diagnostic line available to us is the Ge~I 3039.07\,\AA\ line, which is close to the atmospheric cut-off.\smallskip
    \item{\it Hafnium (Z\,$=$\,72):} 
    Some metal-poor stars display enhanced actinide abundances compared to the expected r-process distribution (`actinide-boost' stars). This is usually traced via the Th/Eu ratio, but Th/Hf is also responsive to the initial conditions, in particular the electron fraction \cite{eichler19}. There are two Hf~II lines (3399.79, 3719.28\,\AA) that are well within the CUBES range, so we investigated their detectability.\smallskip
    \item{\it Bismuth (Z\,$=$\,83):} The only Bi line available is at 3024.64\,\AA, even closer to the atmospheric cut-off than the Ge line, making it incredibly difficult to detect with a ground-based facility. Moreover, it is just shortwards of the region covered by the bluest standard setting of UVES, and in the case of CS~31082-001 the Bi abundance was determined from {\em HST} observations \cite{barbuy11}.\smallskip
    \item{\it Uranium (Z\,$=$\,92):} U~I 3859.57\,\AA\ is a strongly blended line that requires high spectral resolution and sensitivity to be detected, and has only been detected in six stars so far \cite{holmbeck18}.
\end{itemize}

Determining Bi and U abundances is fundamental to understand the nucleosynthesis of very heavy elements, created by the same mechanism in an r-process event. These elements can therefore provide useful constraints on the production rates of r-process elements. Moreover, Pb and Bi are direct decay products of U and Th. For this reason, the U/Bi and U/Th ratios are frequently used as cosmochronometers \cite{hill02,barbuy11}.

\section{Methods}\label{methods}

To investigate the potential performance of CUBES for studies of heavy elements in metal-poor stars we first synthesized a model spectrum for CS~31082-001 using the {\tt TURBOSPECTRUM} code \cite{alvarez98,plez12} and adopting published physical parameters of T$_{\rm eff}$\,$=$\,4825\,K, log\,$g$\,$=$\,1.5 and [Fe/H]\,$=$\,$-$2.9 (from \cite{hill02}). We smoothed and rebinned the model spectrum to match the lowest resolution and sampling of the CUBES design, i.e. $R$\,$\sim$\,22,000 and a full-width half maximum (FWHM) of $\sim$0.14\,\AA\ (sampled by 2.3 pixels). Using this model instead of an observed spectrum in our tests gave us full control of the noise, stellar parameters, and chemical abundances in our tests. 

\begin{figure}
    \centering
    \includegraphics[width=0.66\linewidth]{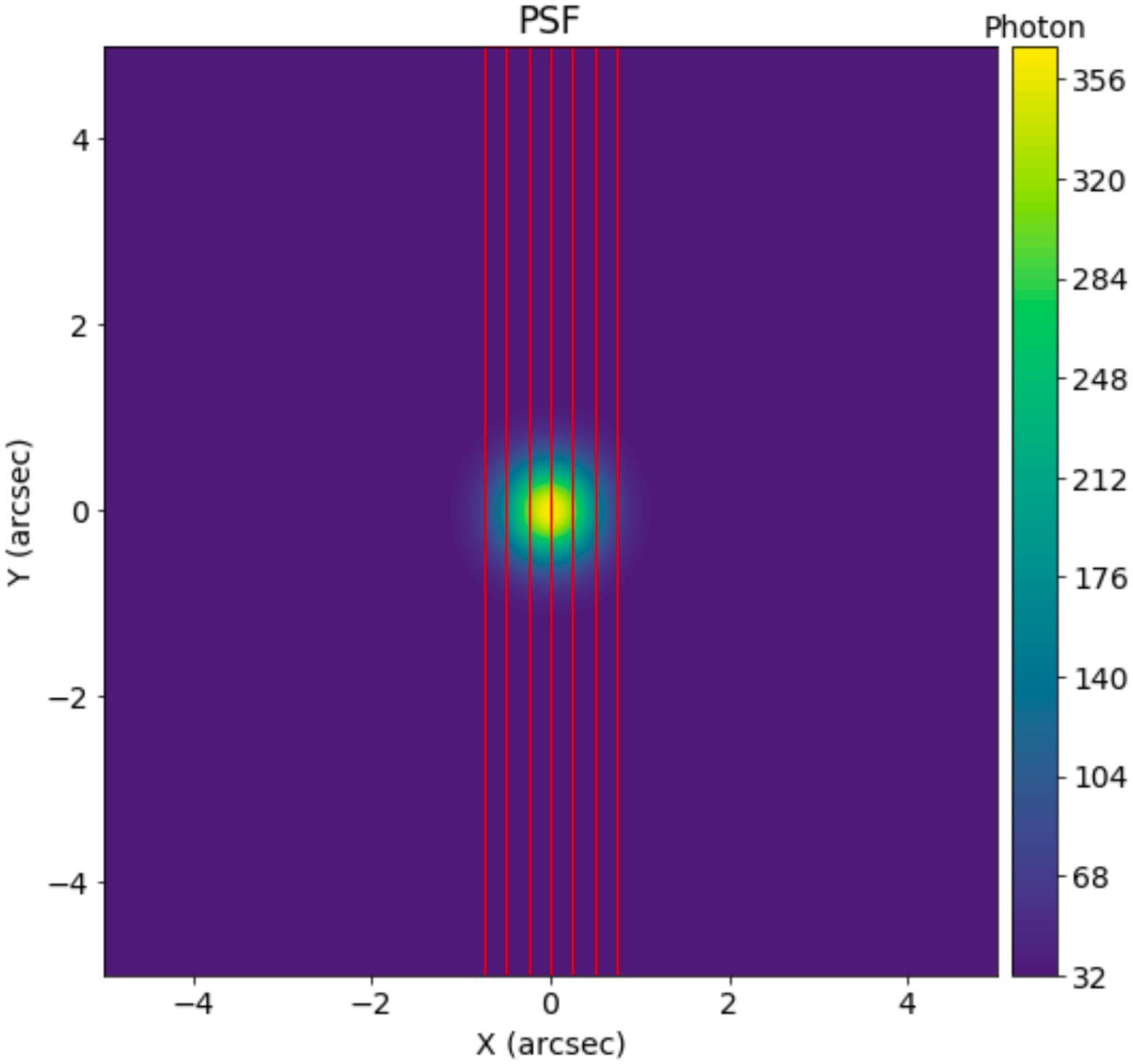}
    \caption{The 0.87$''$ seeing point-spread function (PSF) used in the end-to-end simulation, with the six slices of the image slicer overlaid (that provide a narrower slit to enable greater spectral resolution).}
    \label{psf}
\end{figure}

To mimic a CUBES observation we used this template as our input to the CUBES end-to-end (E2E) simulator that is provided in a Jupyter-notebook \cite{genoni22}. The E2E simulator uses a model point-spread function (PSF) to generate simulated science images; an example of one of these for the default seeing value of 0.87$''$ (at 550\,nm) is shown in Fig.~\ref{psf}. The PSF is then sliced into six narrower slices to deliver the required spectral resolving power \cite{calcines22}. The sky spectrum, atmospheric extinction models, instrument throughput and detector characteristics are then applied to the source spectra of each slice to simulate the appearance of the extracted and co-added data. 


We undertook detailed tests for the Ge~I and Hf~II lines.  To evaluate the detectability of these lines we first synthesised models for the published abundances, log$_{\epsilon}$(Ge)\,$=$\,+0.10 \cite{siqueira-mello13} and log$_{\epsilon}$(Hf)\,$=$\,$-$0.59 \cite{hill02}.  We also synthesised models with larger abundances (+0.35 and +0.65 for Ge, and $-$0.30 and 0.00 for Hf) to investigate the sensitivity of the recovered abundances from the CUBES simulations to the input values. These models were used as inputs for the E2E simulator, adopting the default values of airmass\,$=$\,1.16 and seeing\,$=$\,0.87$''$, for six different exposure times (ranging from 5\,min to 4\,hrs) and for target magnitudes of $U_{\rm E2E}$\,$=$\,16 and 18\,mag; for comparison, CS~31082-001 has $U$\,$=$\,12.5\,mag. However, note that the models are continuum normalised (i.e. have a flat spectral slope), which has consequences on the interpretation of the results, as discussed in Section~\ref{snr_results}.

We then fit the simulated CUBES observations with the reference model spectrum (smoothed and rebinned to the same sampling as CUBES), while varying the abundance of the chosen element in the model. The adopted abundance for each simulated observation is the value that minimises the $\chi^{2}$-statistic when compared to the synthetic spectra. Initial models were fit varying the abundance of the selected element in steps of 0.1\,dex. Once a first minimum in $\chi^{2}$ was found, a finer grid with abundance variations of 0.01\,dex was calculated to arrive at the final estimate. Results from the Ge and Hf tests are summarised in the next section. Similar, illustrative tests for Bi and U are presented briefly in Sect.~5. 

\subsection{Expected uncertainties on measured equivalent widths}

Given the resolution and pixel sampling of CUBES, we used the arguments from
\cite{cayrel88} to estimate the limiting equivalent width ($W$) that could be obtained from high S/N ($>$200) observations. Defining the spectral line as a
Gaussian profile and assuming the spectrum is correctly calibrated (corrected for bias and dark levels, flat-field correction), then: 

\begin{equation}
    W =  \int_{-\infty}^{+\infty} r(x)\,dx = \sum_{i=i_{1}}^{i_{2}} r_{i} \delta x = \delta x \sum_{i=i_{1}}^{i_{2}} r_{i},
\end{equation}
where $r_{i}$ is the flux array and $\delta x $ the pixel size. The uncertainty in the measured widths is then (from \cite{cayrel04}):

\begin{equation}
\label{EW}
      \sigma_w  = \frac{1.5}{S/N} \sqrt{FWHM \times \delta x}.
\end{equation}


Considering the CUBES FWHM of 0.14\,\AA\ and $\delta x$\,$=$\,0.06\,\AA/pixel, a S/N of 200 would yield $\sigma_w$\,$=$\,0.7\,m\AA. However, the above formula neglects the uncertainty on the continuum placement (which can be particularly challenging at these short wavelengths), and the expected uncertainty is usually a factor of two or three larger (see \cite{cayrel04}), i.e. an uncertainty of $\gtrsim$1.5,m\AA\ for CUBES (assuming excellent S/N).

Example models of the Ge~I 3039\,\AA\ and Hf~II 3400\,\AA\ lines and their measured equivalent widths are shown in the left-hand panels of Fig.~\ref{EW}. Alongside them are similar plots for simulated observations of 30 and 240\,min, with the difference between the measured (`observed') width and that from the model indicated. Note that the uncertainties for the longest exposures are already approaching the expected limit. Absolute values of the differential widths for all three lines are given in Table~\ref{EWtab}. 

\begin{figure}
    \centering
    \includegraphics[width=0.9\linewidth]{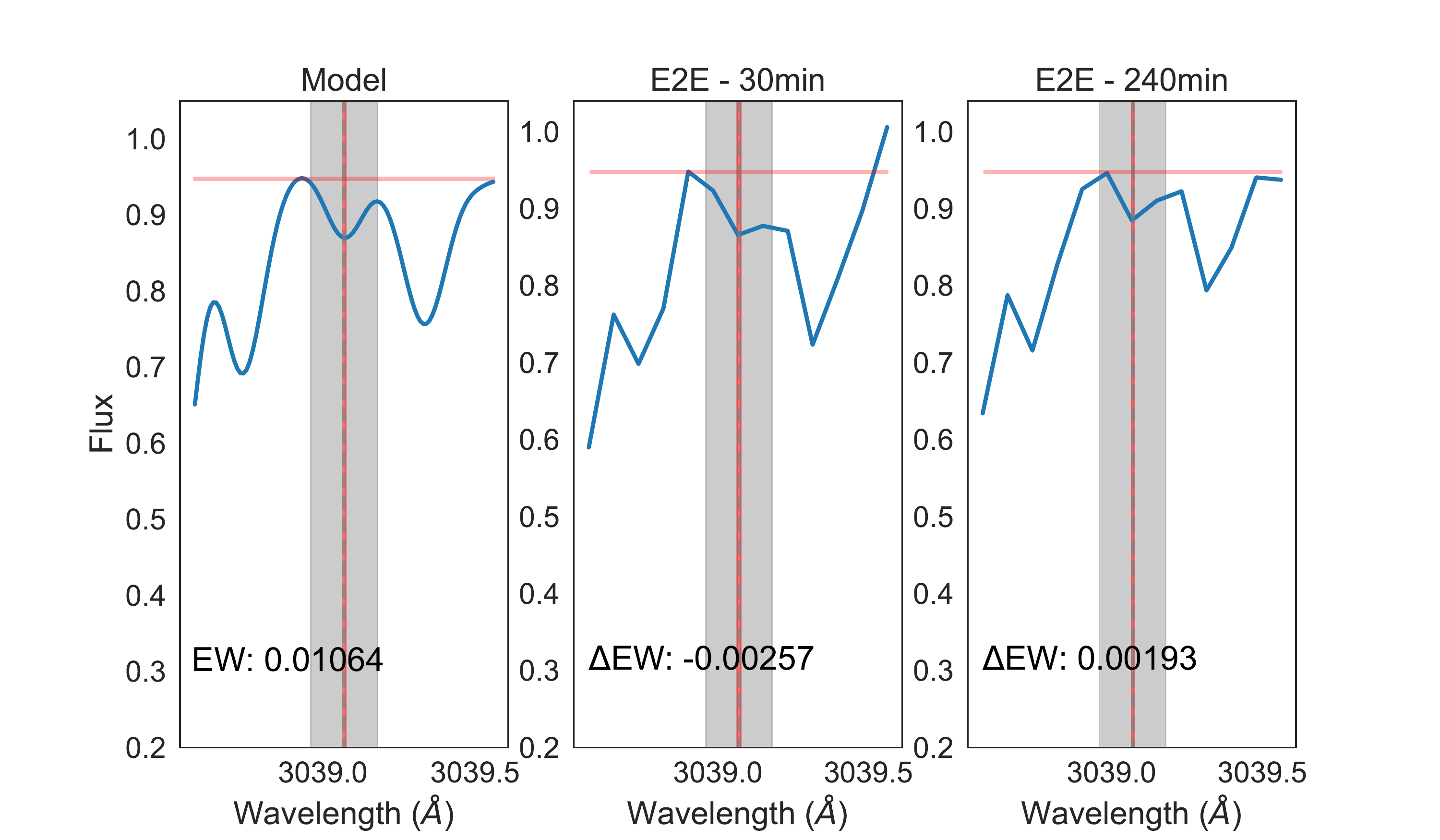}
    \includegraphics[width=0.9\linewidth]{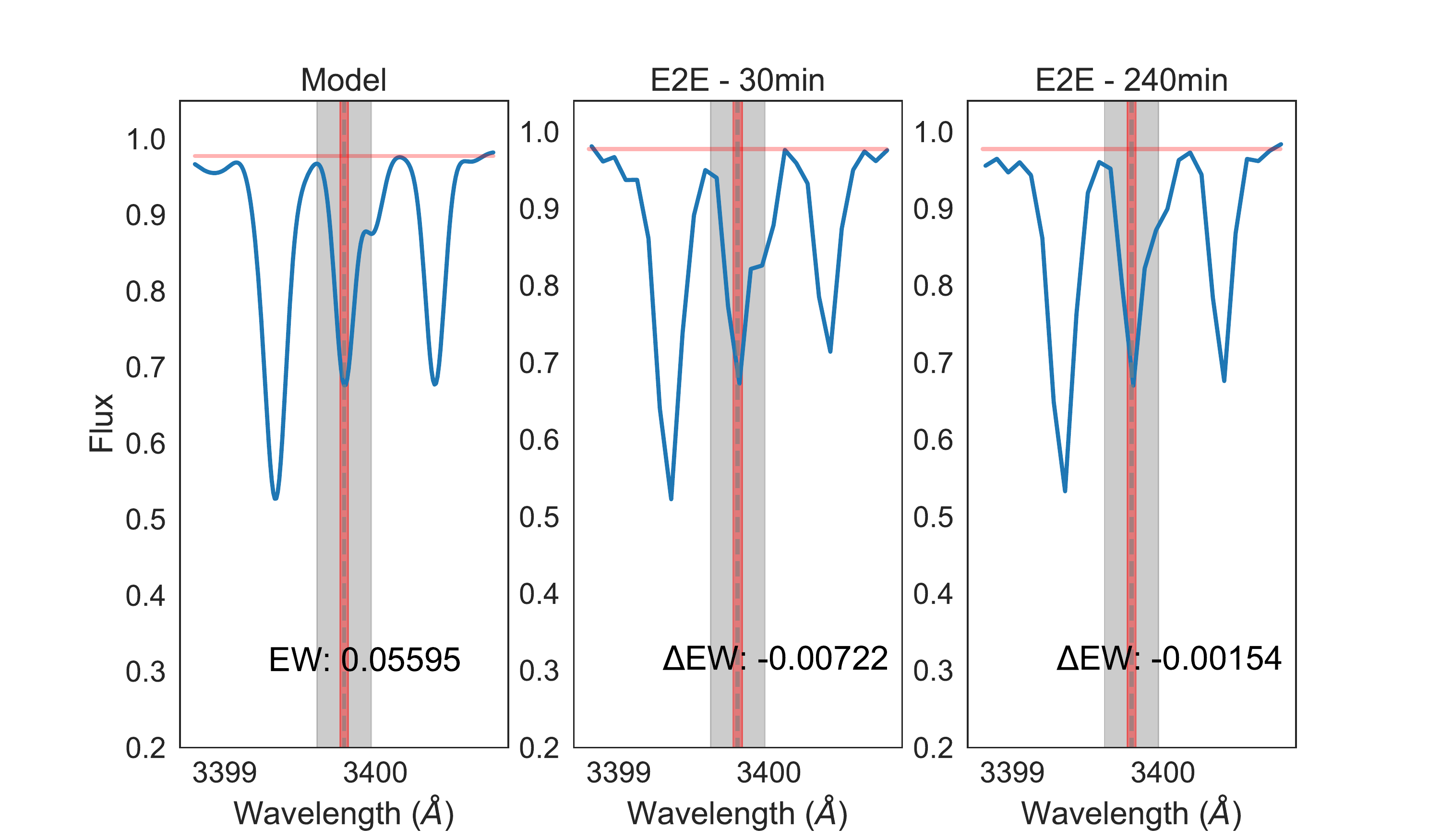}
    \caption{Equivalent widths of Ge~I 3039\,\AA\ (upper panels) and Hf~II 3400\,\AA\ (lower panels) for an input stellar model in the left-hand panels. The centre and right-hand panels show the output spectra from the end-to-end (E2E) simulator for $U$\,$=$\,16\,mag. for 30 and 240\,min, respectively. Differences between the measured width from the simulations compared to the models are given by $\Delta$EW. The regions considered for the estimates are shown in grey.}\label{EW}
\end{figure}


\begin{table}
\caption{Differential equivalent widths ($\Delta$EW) for simulated CUBES observations of the Ge~I 3039.07\,\AA\ and Hf~II 3399.79, 3719.28\,\AA\ lines compared to the published values for CS~31082-001 (used here as the input model abundances).}\label{EWtab} 
\centering                  
\begin{tabular}{c|cccc} 
\hline            
$t_{\rm exp}$  &  \multicolumn{3}{c}{$|\Delta$EW$|$ (m\,\AA)} & \\ 
(min)          & Ge~I 3039\,\AA & Hf~II 3400\,\AA & Hf~II 3719\,\AA & \\
\hline  
30  & 2.57  & 7.22  & 6.33 \\
60  & 3.35  & 5.46  & 1.67 \\
120 & 3.16  & 1.67  & 4.38 \\
240 & 1.93  & 1.54  & 2.22 \\
\hline
\end{tabular}
\end{table}

\section{Results: Ge \& Hf}
\label{results}

\subsection{Continuum signal-to-noise}\label{snr_results}
We first estimated the continuum S/N (per pixel) for the simulated spectra in regions close to the Ge~I 3039\,\AA\ and Hf~II 3719\,\AA\ lines for insight into the exposure times required to recover robust abundances. These estimates (for the initial abundance models, i.e. A(Ge)\,$=$\,+0.10 and A(Hf)\,$=$\,$-$0.59) are summarised in Table~\ref{snrtab}.

Caution is required in interpreting the S/N estimates from the E2E simulations at the bluewards end of the CUBES range as we have (necessarily) used continuum-normalised model spectra which have been scaled to the desired $U$-band magnitudes. The Johnson $U$-band filter has an effective central wavelength of 366\,nm, with a width of 65\,nm (e.g. \cite{bessell2005}). Thus, when our model spectra were scaled to $U$\,$=$\,16 and 18\,mag., the resulting model flux will be a good enough approximation to the real astrophysical flux for the Hf~II 3719\,\AA\ line as it is close to the centre of the photometric band (i.e. the estimated S/N value is realistic for the true flux distribution of a star such as CS~31082-001). 

However, this is not the case for the Ge~I 3039\,\AA\ line, which is located in the bluewards tail of the $U$-band filter curve. Given its low effective temperature (T$_{\rm eff}$\,$=$\,4825\,K), the flux distribution of CS~31082-001 declines rapidly towards shorter wavelengths across the $U$-band filter. This means that use of continuum-normalised models as inputs to the E2E simulator give an unrealistically large estimate of the flux at short wavelengths when they are scaled to the desired $U$-band magnitude.  This does not impact on the ability to recover the targeted abundance as a function of the S/N from the E2E simulations, but means that the realistic observed magnitude for which this level of performance can be achieved needs to be quantified. In summary, the effective $U$-band magnitude ($U_{\rm eff}$) for the appropriate flux level for a star similar to CS~31082-001 at 3039\,\AA\ will be brighter than the values used in the E2E simulations.

To estimate of the scale of this effect, we used the CUBES Exposure Time Calculator (ETC) \cite{genoni22}, adopting the same conditions as the E2E simulations (airmass\,$=$\,1.16, seeing\,$=$\,0.87$''$) and using one of the standard template spectra (T$_{\rm eff}$\,$=$\,4500\,K, log($g$)\,$=$\,1.5, [M/H]\,$=$\,$-$3.5), that is a reasonable match to the parameters of CS~31082-001.
The S/N predicted by the ETC matches that from the E2E by adopting $\Delta U$\,$\sim$\,$-$1.75\,mag. That is, using the T$_{\rm eff}$\,$=$\,4500\,K template, the S/N predicted by the ETC for $U_{\rm ETC}$\,$=$\,14.25 matches the E2E result for $U_{\rm E2E}$\,$=$\,16 (and similarly for $U_{\rm ETC}$\,$=$\,16.25 for $U_{\rm E2E}$\,$=$\,18).  We emphasise that this is due to the difference in the spectral slope of the models at the short wavelengths, and that the E2E and ETC results at 3730\,\AA\ are in good agreement.


\begin{table}
\caption{Predicted continuum signal-to-noise (S/N) per pixel from the end-to-end (E2E) CUBES simulations near the Ge~I 3039\,\AA\ line and the redder Hf~II line (3719\,\AA). The adopted magnitudes for the E2E simulations were $U_{\rm E2E}$\,$=$\,16 and 18 mag. For simulations of CS~31082-001 these values are a good approximation to the real flux at 3730\,\AA, but given the use of continuum-normalised model spectra, the effective observed magnitudes ($U_{\rm eff}$) for the Ge~I 3039\,\AA\ line are $\sim$1.75\,mag brighter (see Section~\ref{snr_results} for discussion).
}\label{snrtab} 
\centering                  
\begin{tabular}{c|cc|cc} 
\hline            
$t_{\rm exp}$ &  S/N$_{\lambda 3050}$ & S/N$_{\lambda 3730}$ & S/N$_{\lambda 3050}$ & S/N$_{\lambda 3730}$ \\
(min) & ($U_{\rm eff}$\,$\sim$\,14.25) & ($U_{\rm E2E}$\,$=$\,16) & ($U_{\rm eff}$\,$\sim$\,16.25) & ($U_{\rm E2E}$\,$=$\,18) \\ 
\hline  
5 & 7   & 24  & 1 & 7 \\
10 & 12  & 34  & 3 & 11 \\
30 & 25  & 65  & 6 & 23 \\
60 & 36  & 92  & 10 & 37 \\
120 & 53  & 148 & 15 & 47 \\
240 & 89  & 190 & 20 & 70 \\
\hline
\end{tabular}
\end{table}

\subsection{Abundance estimates}
\begin{figure}
\begin{center}
\begin{minipage}{.5\textwidth}
    \includegraphics[width=1.0\linewidth]{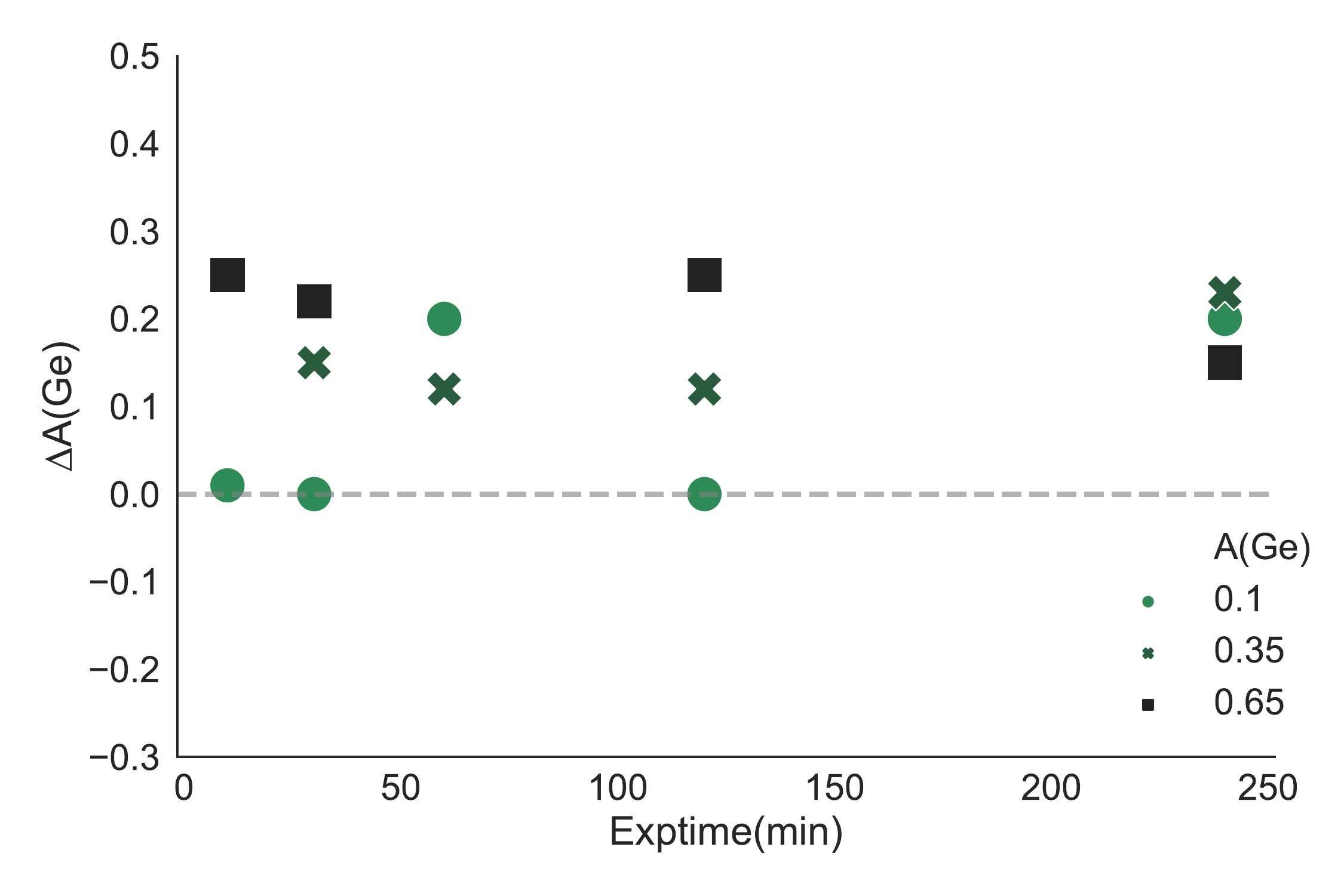}
\end{minipage}%
\begin{minipage}{.5\textwidth}
    \includegraphics[width=1.0\linewidth]{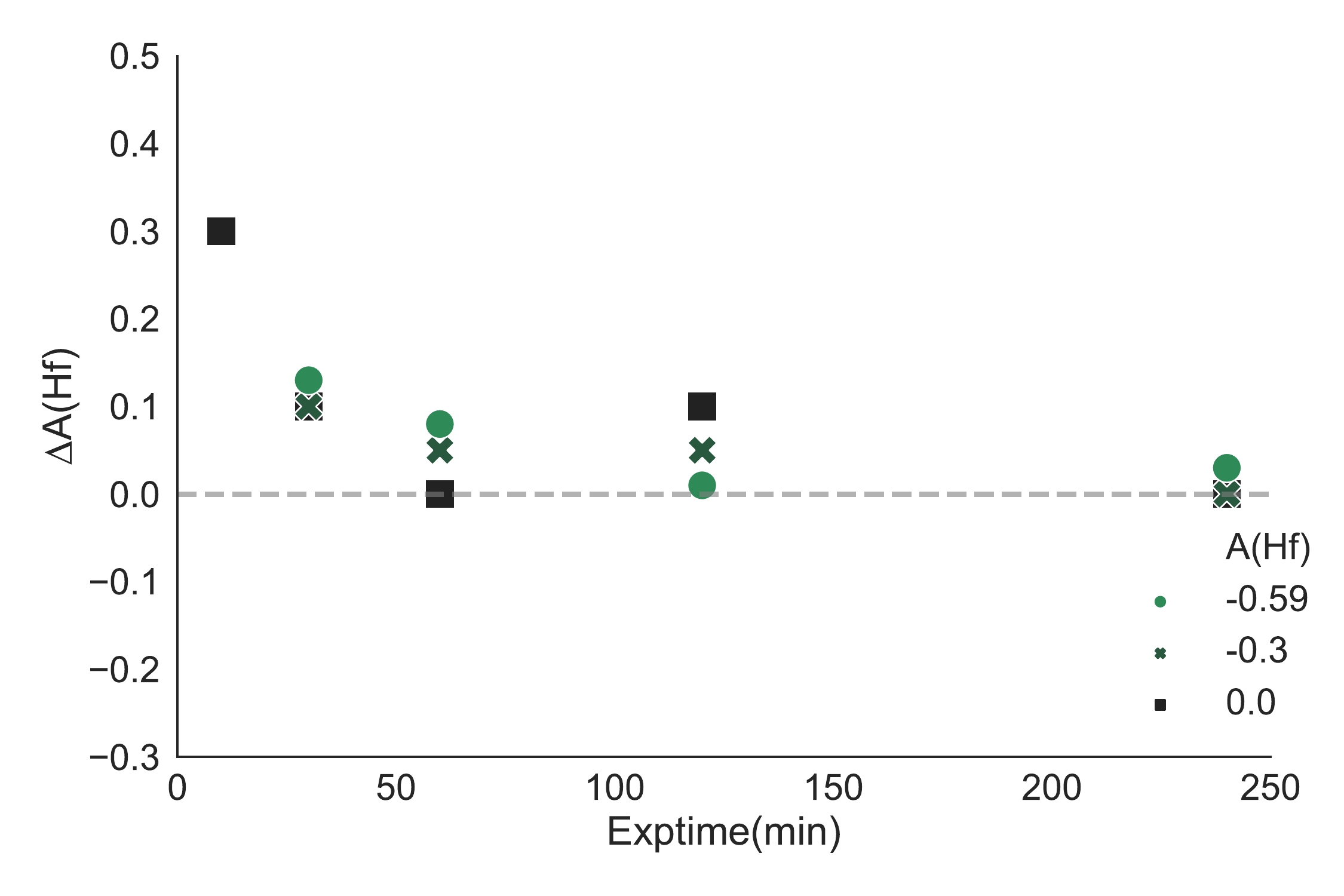}
\end{minipage}
\end{center}
\caption{Abundance uncertainties ($\Delta$A) compared to the input values for the $U_{\rm E2E}$\,$=$\,16\,mag. CS31082-001 simulations. {\it Left:} Ge~I 3039\,\AA, {\it Right:} Hf~II 3400.}\label{U16tests}


\begin{center}
\begin{minipage}{.5\textwidth}
  \includegraphics[width=1.0\linewidth]{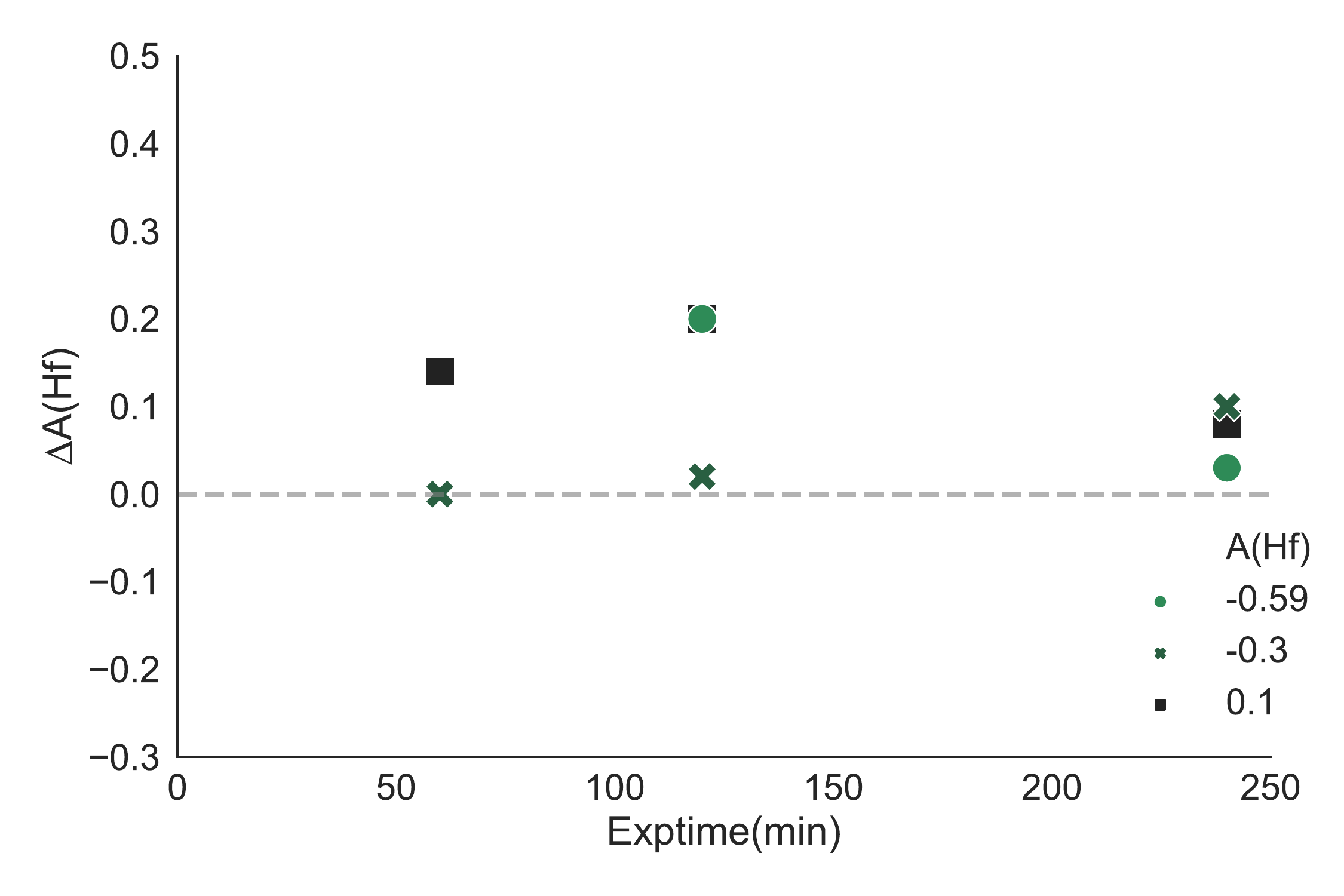}
\end{minipage}
\end{center}
  \caption{Abundance uncertainties ($\Delta$A) compared to the input values for the $U_{\rm E2E}$\,$=$\,18\,mag. CS31082-001 simulations for Hf~II 3400.}\label{U18tests}
\end{figure}

Differential abundances, $\Delta$A(X), between the input values and those estimated from the simulated observations with $U_{\rm E2E}$\,$=$\,16\,mag. for Ge and Hf are shown in Fig.~\ref{U16tests}.
For observations of $>$1\,hr, the uncertainty on the Hf abundance is of order $\pm$0.1\,dex, with an even smaller dispersion for the longest (4\,hr) integration. Compared to the S/N estimates from Table~\ref{snrtab}, these results argue that with S/N\,$\gtrsim$\,100 we can recover estimates of Hf abundances to $\pm$0.1\,dex. 

Given the lower S/N at the shorter wavelengths, the dispersion on the Ge results is larger for shorter exposures, but with reasonable {\it differential} agreement (to $\pm$0.1\,dex) for the longest simulated exposure (with S/N\,$=$\,89 from Table~\ref{snrtab}). However, we note the systematic offset to larger estimated abundances for the Ge results (and, to a lesser extent, Hf too). This is not surprising for a weak line such as Ge~I 3039\,\AA, and is probably a consequence of the challenges of continuum normalisation in these rich spectral regions, where the real continuum level can be underestimated due to the pseudo-continuum from the line blends (thus requiring an enhanced abundance value to reproduce a given line feature). Nonetheless, for the current purpose, our tests suggest that differential Ge abundances (to $\pm$0.1\,dex) should be possible for a star similar to CS~31082-001 with $U$\,$\sim$\,14.25\,mag. within a few hours.

Results for the simulations with $U_{\rm E2E}$\,$=$\,18\,mag. are shown in Fig.~\ref{U18tests}.  Mirroring the results in terms of the S/N estimates for $U_{\rm E2E}$\,$=$\,16\,mag., the Hf abundance can be recovered to better than $\pm$0.1\,dex for the 4\,hr exposures (with S/N\,$=$\,70). The Ge abundance is more challenging for such a faint target as we are limited to S/N\,$=$\,20 even in a 4\,hr observation.  Nonetheless, even at such modest S/N (see results for 30\,min in Fig.~\ref{U16tests}) we can potentially constrain the Ge abundance to $\sim$0.2-0.3\,dex. Achieving even this level of precision for such faint stars would be remarkable compared to our current capabilities.

\begin{figure}
    \centering
    \includegraphics[width=0.8\linewidth]{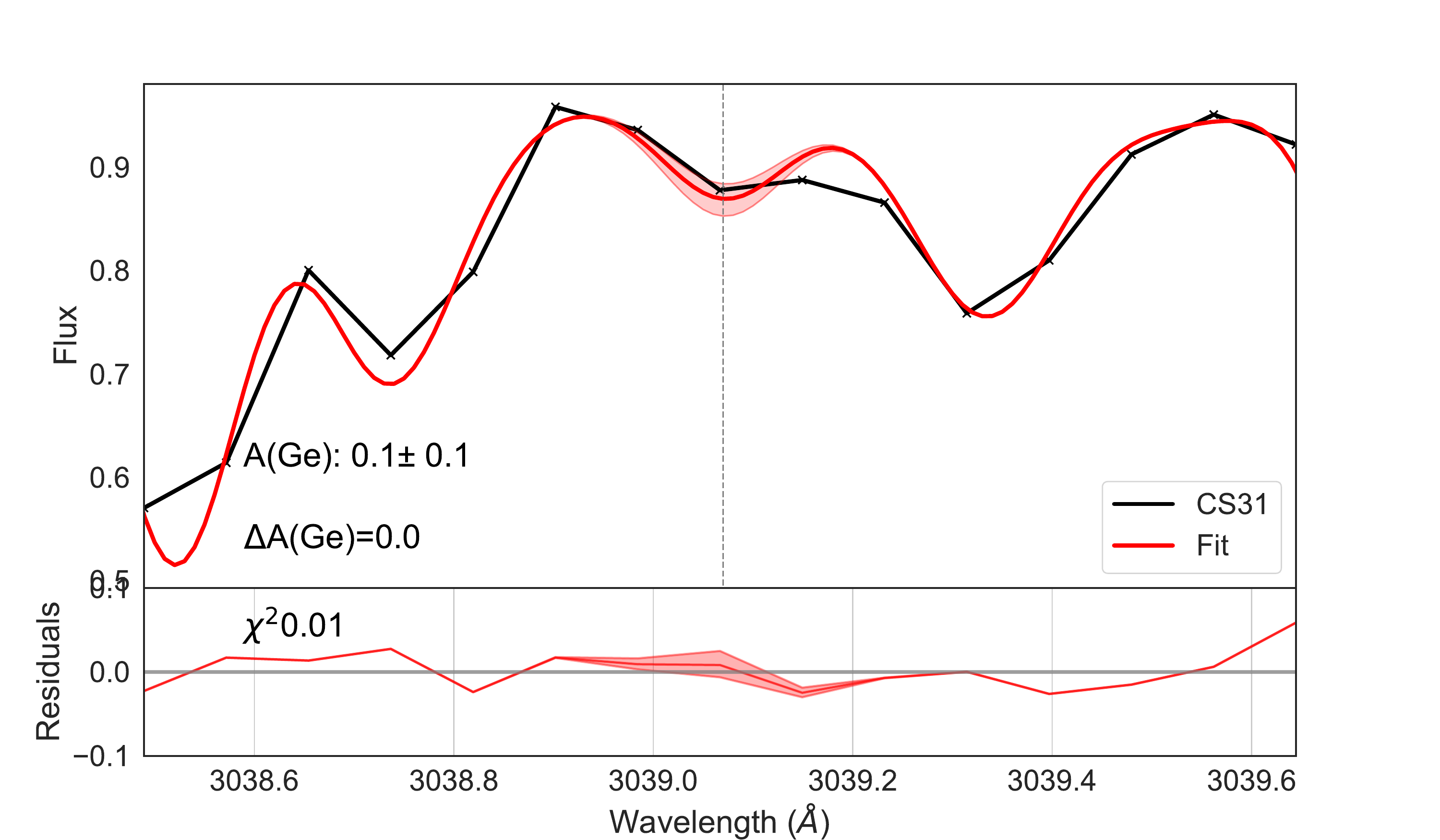}
    \caption{Simulated 2\,hr CUBES spectrum of CS~31082-001 (with $U_{\rm E2E}$\,$\sim$\,16\,mag.) for the  Ge~I 3039\,\AA\ region. The reference model (A(Ge)\,$=$\,$+$0.10) is plotted in black, with the best fit model (A(Ge)\,$=$\,$+$0.1) in red. The red shaded area indicates $\pm$\,0.1\,dex in Ge abundance.}
    \label{plotsGe}

    \centering
    \includegraphics[width=0.8\linewidth]{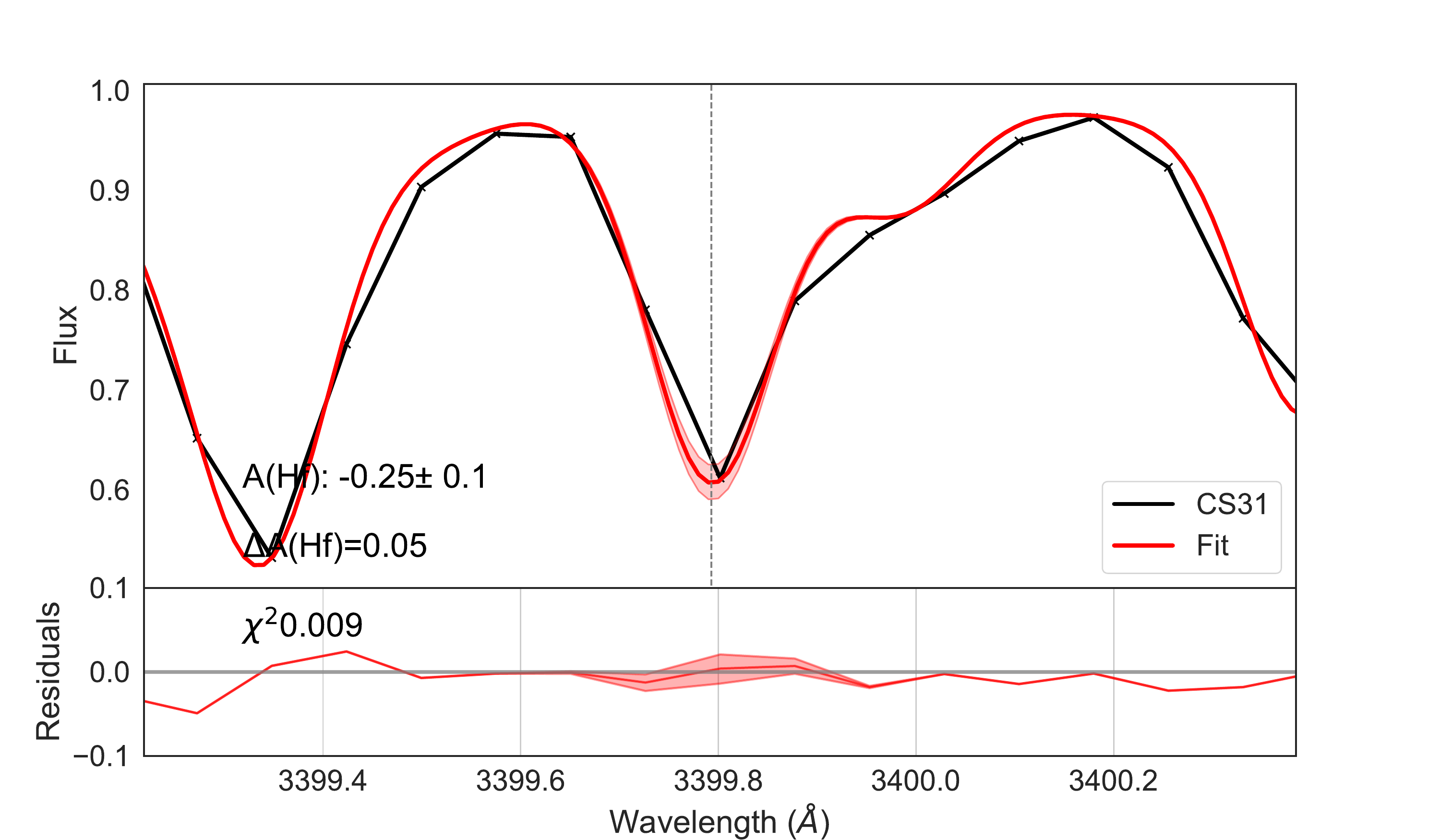}
    \caption{Simulated 2\,hr CUBES spectrum of CS~31082-001 (with $U_{\rm E2E}$\,$=$\,16\,mag.) for the  Hf~II 3400\,\AA\ region. The reference model (A(Hf)\,$=$\,$-$0.30) is plotted in black, with the best fit model (A(Hf)\,$=$\,$-$0.25) in red. The red shaded area indicates $\pm$\,0.1\,dex in Hf abundance.}
    \label{plotsHf1}

    \centering
    \includegraphics[width=0.8\linewidth]{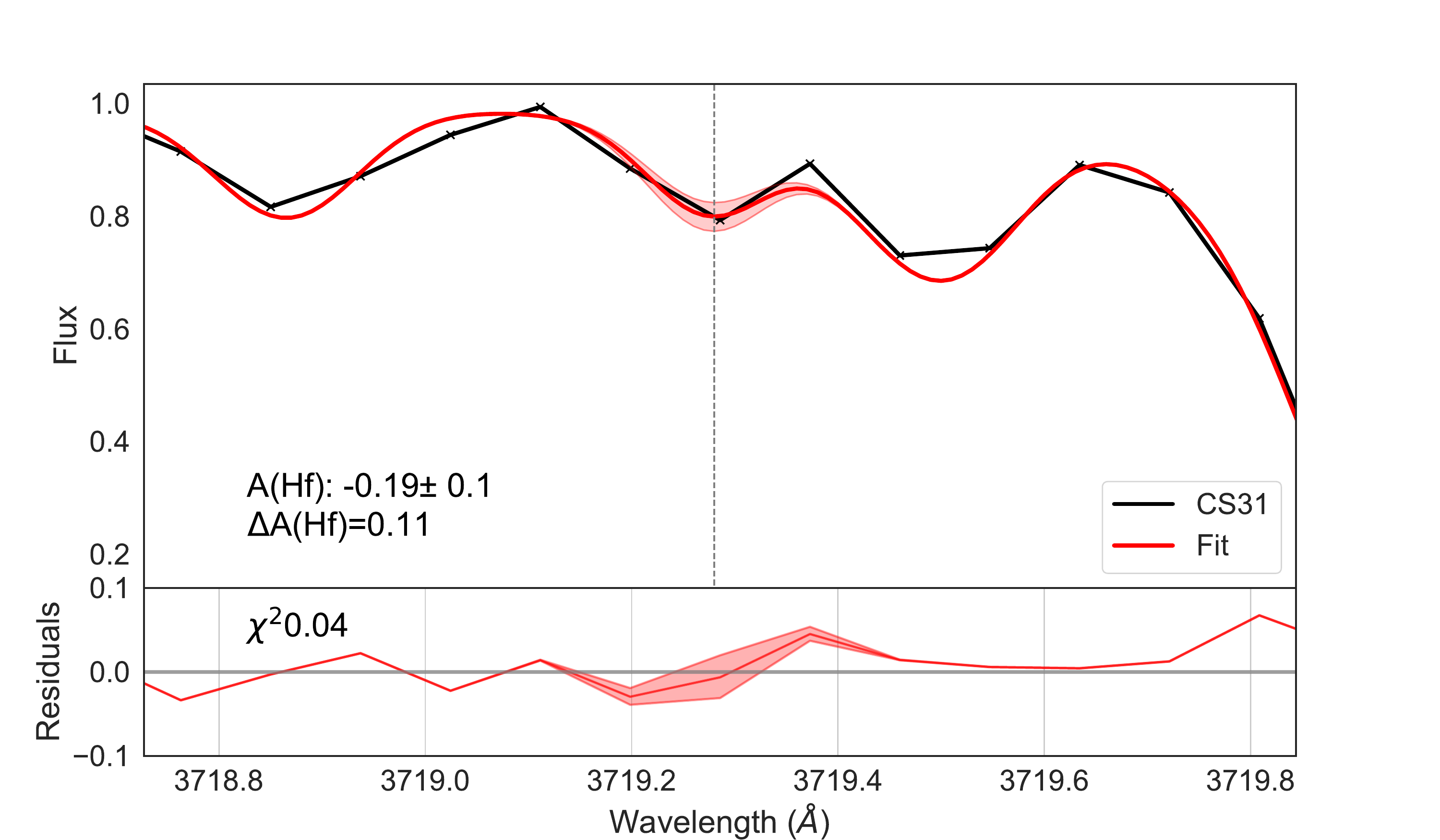}
    \caption{Simulated 2\,hr CUBES spectrum of CS~31082-001 (with $U_{\rm E2E}$\,$=$\,16\,mag.) for the  Hf~II 3719\,\AA\ region. The reference model (A(Hf)\,$=$\,$-$0.30) is plotted in black, with the best fit model (A(Hf)\,$=$\,$-$0.19) in red. The red shaded area indicates $\pm$\,0.1\,dex in Hf abundance.}
    \label{plotsHf2}
\end{figure}

Example fits for the $U_{\rm E2E}$\,$=$\,16\,mag. simulations are shown in Figs.~\ref{plotsGe}, \ref{plotsHf1} and \ref{plotsHf2}, in which the simulated CUBES data (in black) and model spectra (in red, prior to binning for display purposes) have been continuum normalised.

\section{Results: Bi \& U}

\begin{figure}
    \centering
    \includegraphics[trim=0cm 2cm 0cm 0cm, width=0.9\linewidth]{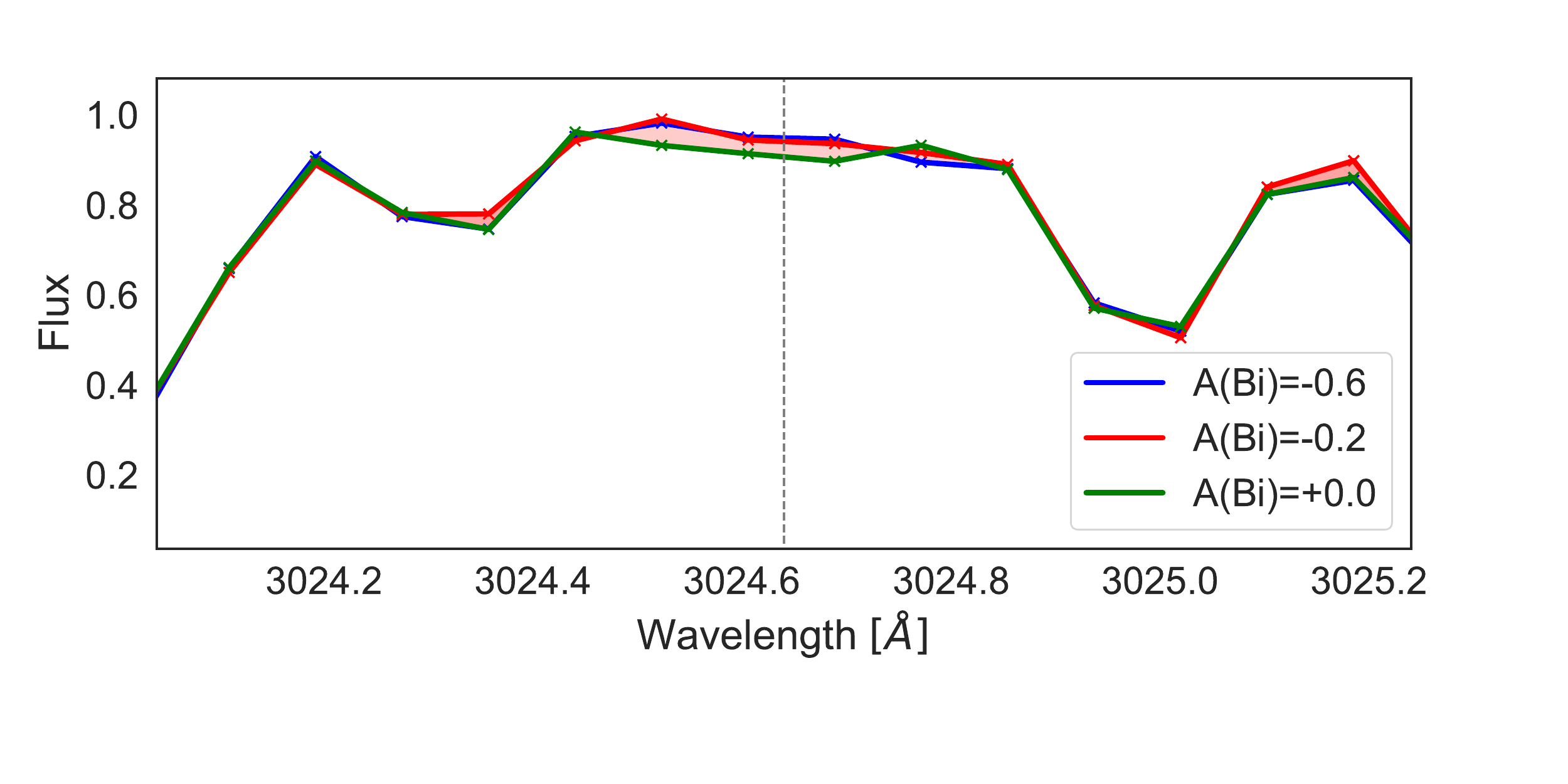}
    \caption{Simulated 4\,hr CUBES observation of the Bi~I 3024.64\,\AA\ line for $U_{\rm E2E}$\,$=$\,16\,mag. and models with A(Bi)\,$=$\, $-$0.6, $-$0.2, and 0.0. The pink shaded regions are to highlight the differences between the models (where both abundance and S/N contribute).}
    \label{Bi}

    \centering
    \includegraphics[trim=0cm 2cm 0cm 0cm, width=0.9\linewidth]{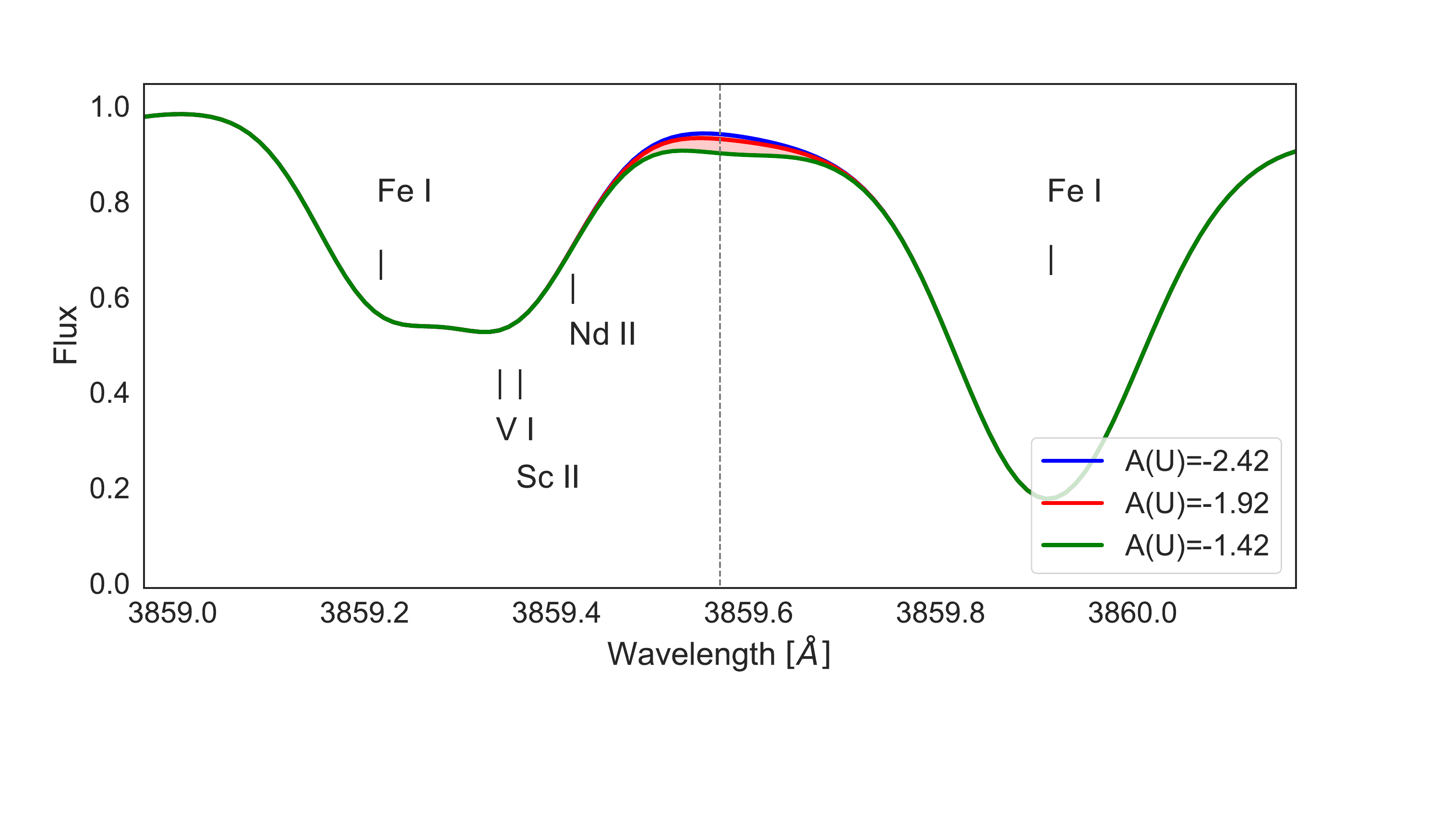}
    \includegraphics[trim=0cm 2cm 0cm 0cm, width=0.9\linewidth]{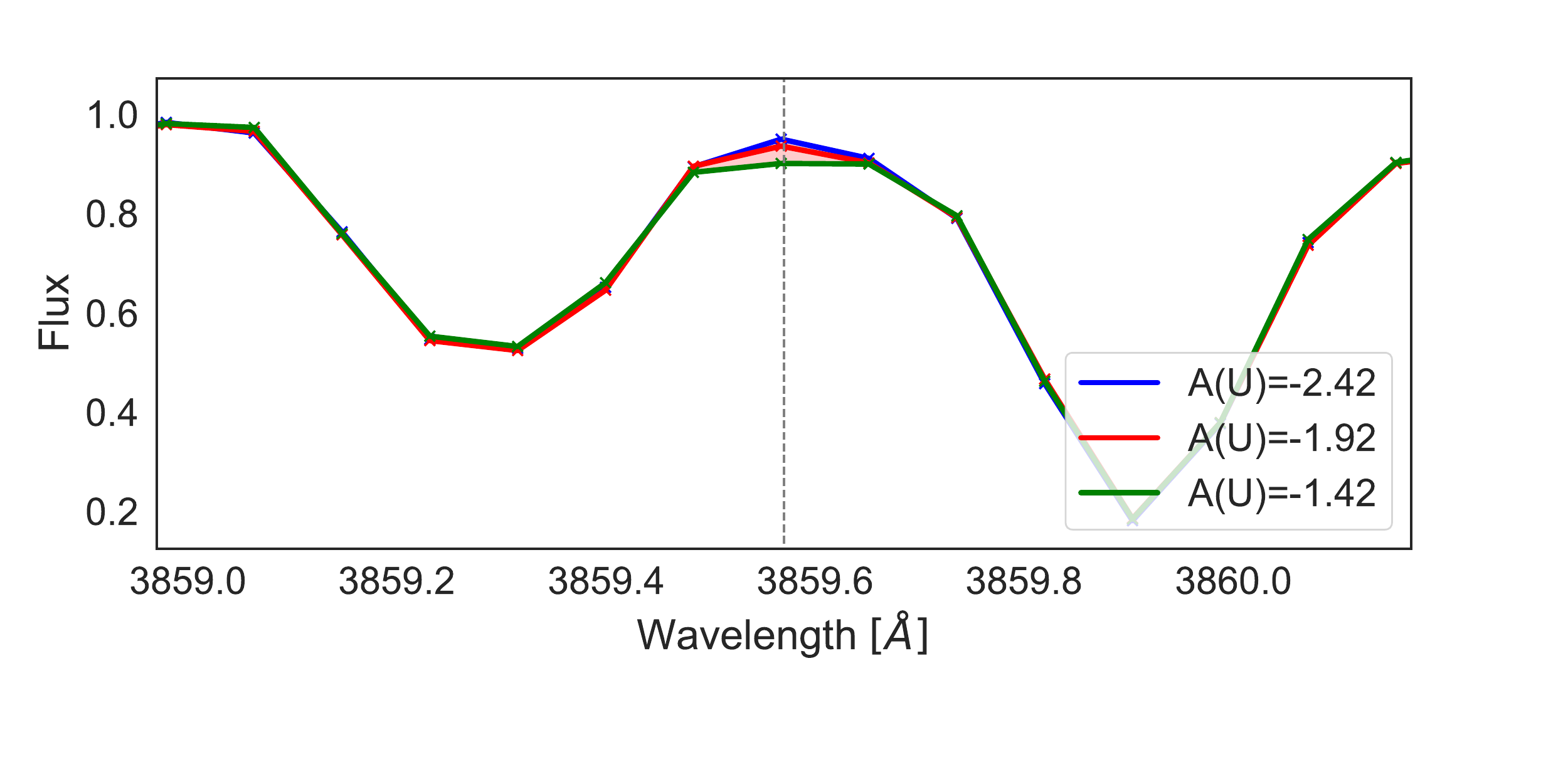}
    \caption{{\it Upper panel:} Model spectra in the region of U~II 3859.57\,\AA\ for $U_{\rm E2E}$\,$=$\,16\,mag. and with A(U) as indicated in the legend. The spectra have been convolved to $R$\,$=$\,22,000 but are left oversampled to better highlight the other contributing lines. {\it Lower panel:} Simulated 4\,hr CUBES observation of the same region. As in Fig.~\ref{Bi}, the shaded regions highlight the differences between the models.}
    \label{U}
\end{figure}

We also used the E2E software to simulate CUBES observations of the Bi~I 3024.64\,\AA\ and U~II 3859.57\,\AA\ lines, in which we varied the model abundances around the published values of A(Bi)\,$=$\,$-$0.4 \cite{barbuy11} and A(U)\,$=$\,$-$1.92 \cite{hill02}.  Simulated spectra from 4\,hr exposures for $U_{\rm E2E}$\,$=$\,16\,mag. for the Bi and U regions are shown in Figs.~\ref{Bi} and \ref{U}, respectively. Given the weakness of these lines combined with the impact of nearby blends and difficulty of continuum normalisation, precise abundance determinations are particularly challenging for these elements. To emphasise this in the case of the weak U~II line, the upper panel of Fig.~\ref{U} shows the input models and identifies the stronger, nearby lines that influence the final appearance of the U~II line in the CUBES simulations\footnote{Nd~II 3859.413\,\AA, V~I 3859.337\,\AA, Sc~II 3859.358\,\AA, and Fe~I 3859.213, 3859.911\,\AA\, as well as a contribution from a CN feature.}. 

Nonetheless, given good S/N ($>$100), CUBES observations of both the Bi and U lines should be able to place initial constraints on their abundances, even if just upper limits in the case the lower abundances considered in the examples here.



\section{Concluding remarks}

We have presented analysis of simulated CUBES observations of faint ($U_{\rm E2E}$\,$=$ 16 and 18\,mag.) metal-poor stars, using the well-studied CS~31082-001 star as a template.  We selected two elements (Ge and Hf) with relatively isolated lines to investigate the performance of CUBES near the blueward limit of the instrument (Ge) and at longer wavelengths (Hf) in detailed abundance studies of metal-poor stars.  We also investigated the weak Bi~I and U~II lines, which are two of the most challenging lines for abundance estimates with current facilities. Our key findings are:

\begin{itemize}
    \item {Ge: From a 4\,hr integration for $U_{\rm E2E}$\,$=$\,16\,mag. we recover sufficient S/N to obtain differential Ge abundances to $\sim\pm$0.1\,dex. Given the discussion in Section~\ref{snr_results}, this translates to an effective $U$-band magnitude for a star similar to CS~31082-001 of $U$\,$\sim$\,14.25\,mag.}\smallskip
    \item {Hf: Given the better transmission of the atmosphere at longer wavelengths, we can recover Hf abundances (from the Hf~II 3400, 3719\,\AA\ lines) to $\sim$0.1\,dex in just 1\,hr for $U$\,$=$\,16\,mag., and in 4\,hr for $U$\,$=$\,18\,mag.}\smallskip
    \item{Bi and U: Our simulated 4\,hr observations indicate we should be able to place initial constraints on the abundances (even if just upper limits) of these more challenging, weak lines.} 
\end{itemize}

A critical factor in arriving at accurate (rather than precise) abundances in the CUBES range is the ability to correctly estimate the continuum level when calibrating and analysing the spectra (as evidenced by the systematic offset in the Ge results in Fig.~\ref{U16tests}, and even more critical for the weak Bi and U lines). As such, quantitative abundance studies of metal-poor stars with CUBES will require the physical properties of future targets (temperatures, gravities, metallicity, microturbulence etc) to be well constrained from other observations, which could include simultaneous observations with the fibrelink to UVES that was included in the Phase~A design \cite{zanutta22}. 

In summary, by combining unprecedented near-UV sensitivity with sufficient spectral resolution for quantitative stellar studies \cite{ernandes20}, CUBES will provide an exciting and unique capability for studies of chemical abundances and stellar nucleosynthesis. The simulations presented here for a faint analogue of CS~31082-001 have demonstrated the high-precision, differential abundances will be possible for both Ge and Hf for targets that are two-to-three magnitudes fainter than possible with current facilities. In terms of the $U$-band magnitude achievable, the predicted performance at $<$350\,nm is strongly dependent on the spectral energy distribution of the target. The temperature of CS~31082-001 is quite cool compared to those of the stars that will comprise the majority of future CUBES targets, so even greater gains can be anticipated for elements such as Ge and Bi in studies of hotter stars.


\begin{acknowledgements}
HE acknowledges a CAPES PhD fellowship, and a PRINT/CAPES fellowship in support of his year of study at the University of Edinburgh. HE and BB acknowledge partial financial support from FAPESP, CNPq, and CAPES - Financial code 001. HE and CJE also acknowledge support from the Global Challenges Research Fund (GCRF) from UK Research and Innovation. 
\end{acknowledgements}

%
\section*{Conflict of interest}

The authors declare that they have no conflict of interest.



\end{document}